\documentclass[reprint,
 amsmath,amssymb,
 aps,
]{revtex4-2}

\usepackage{graphicx}

\graphicspath{ {./figs/} }

\begin{document}

\title{Isochronous Data Link Across a Superconducting Nb Flex\\ Cable
  with 5\,femtojoules per Bit}

\author{
  Haitao Dai\textsuperscript{1},
  Corey Kegerreis\textsuperscript{1},
  Deepal Wehella Gamage\textsuperscript{1},
  Jonathan Egan\textsuperscript{1},
  Max Nielsen\textsuperscript{1},
  Yuan Chen\textsuperscript{1},
  David Tuckerman\textsuperscript{2},
  Sherman E. Peek\textsuperscript{3},
  Bhargav Yelamanchili \textsuperscript{3},
  Michael Hamilton\textsuperscript{3},
  Rabindra N. Das\textsuperscript{4},
  Anna Herr\textsuperscript{1},
  and Quentin Herr\textsuperscript{1}
}
\affiliation{\textsuperscript{1}Northrop Grumman Corp., Baltimore, MD 21240}
\affiliation{\textsuperscript{2}Tuckerman \& Associates, Inc., Lafayette, CA 94549}
\affiliation{\textsuperscript{3}Department of Electrical and Computer Engineering, Auburn University, AL 36849}
\affiliation{\textsuperscript{4}MIT Lincoln Laboratory, Lexington, MA 02420}

\thanks{This research is based upon work supported in part by the
  ODNI, IARPA, via ARO. The views and conclusions contained herein are
  those of the authors and should not be interpreted as necessarily
  representing the official policies or endorsements, either expressed
  or implied, of the ODNI, IARPA, or the U.S. Government.}

\date{\today}

\begin{abstract}
Interconnect properties position superconducting digital circuits to
build large, high performance, power efficient digital systems. We
report a board-to-board communication data link, which is a critical
technological component that has not yet been addressed. Synchronous
communication on chip and between chips mounted on a common board is
enabled by the superconducting resonant clock/power network for RQL
circuits. The data link is extended to board-to-board communication
using isochronous communication, where there is a common frequency
between boards but the relative phase is unknown. Our link uses
over-sampling and configurable delay at the receiver to synchronize to
the local clock phase. A single-bit isochronous data link has been
demonstrated on-chip through a transmission line, and on a multi-chip
module (MCM) through a superconducting tape between driver and
receiver with variable phase offset. Measured results demonstrated
correct functionality with a clock margin of 3\,dB at 3.6\,GHz, and
with 5\,fJ/bit at 4.2\,K.
\end{abstract}

\maketitle

Superconducting digital circuits have the potential to enable digital
systems with very high computation density
\cite{holmes2013energy}. The combination of low power logic
\cite{herr2011ultra} and low loss interconnects
(e.g. \cite{filippov2017experimental} and references thereof) shifts
the design trade-off towards heterogeneous distributed architectures
with multiple chips on multiple boards. Volumetric cooling with
efficient heat transfer allows such a system to be physically small,
with packing density limited only by the physical dimensions of the
connectors on the boards. Superconducting systems trade modest chip
integration density for high packaging density.

Efficient distributed heterogeneous architectures require tremendous
communication bandwidth on-board and board-to-board. Design challenges
involve both the interconnects themselves, and synchronization between
driver and receiver. Nb interconnects having 700\,GHz analog bandwidth
can cover distance with appropriate data encoding and geometry of the
wires \cite{talanov2021propagation}. A bandwidth-efficient driver
producing 10 SFQ pulses per bit, with an analog bandwidth of 35\,GHz
and 0.58\,fJ per bit has been reported in
\cite{egan2021MCM}. This bandwidth-efficient driver has been used to
demonstrate synchronous communication between 10$\times$10\,mm chips
on an MCM with interconnect length up to 54\,mm. Superconducting Nb
flex tape on Polyimide has been proposed for board-to-board
communication \cite{tuckerman2016flexible}, \cite{gupta2019thin}. Tape
enables longer range communication relative to MCM because of very low
dielectric losses, $\tan\delta\approx10^{-4}$, at cryo temperatures
\cite{bai2016cryogenic} and due to relatively large dimensions for the
traces and dielectric thickness. The 35\,Gb/s serial bandwidth of such
tape is five times higher than state-of-the-art CMOS (e.g.,
\cite{miller2017attojoule} and references thereof) at three orders of
magnitude less power dissipation while including a cryocooler overhead
of 300\,W/W.

\begin{figure}[b]
 \centering
 \includegraphics[width=3.4in]{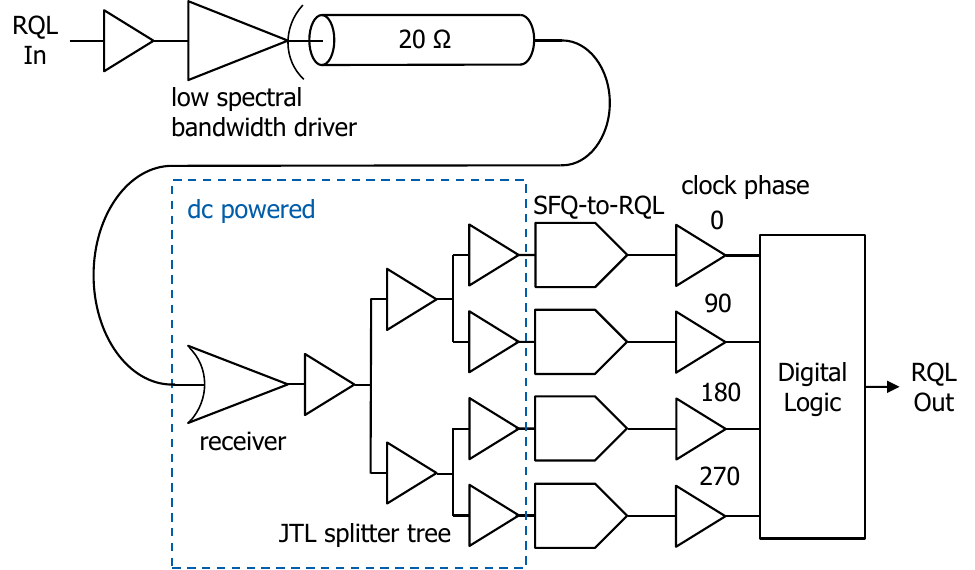}
 \caption{Block diagram of the isochronous communication link with
   driver and receiver operating at the same frequency but with
   unknown static phase offset. The driver converts the ps-scale SFQ
   data encoding into a wider signal to reduce dispersion in the Nb
   wire. The data is received by the DC-biased receiver and is later
   aligned to four phases of AC resonant clock by the digital block.
 \label{block_diagram}}
\end{figure}

Superconducting RQL technology also provides a simple and robust way
to implement synchronization between boards. RQL uses a synchronous
resonator-based clock distributed across chips on the MCM
\cite{strong2021ZOR}. The resonators at 4.2\,K have a high quality factor of
500 that ensures stability, have 30\% efficiency, and are driven by
precise external frequency sources. These features allow RQL to
support isochronous communication between multiple boards, driven at
the same frequency but with an arbitrary phase relationship between
them. The receiver identifies the relative phase between boards
during initialization, and adjusts the data link accordingly. This
scheme is quite simple and lightweight compared to standard SerDes
receivers, without the large overhead in hardware and power
dissipation.

In this paper we present the design of the isochronous receiver,
consisting of 1) an analog circuitry performing oversampling of input
signal using four available phases of RQL clock, and 2) a digital
block performing arbitration among the phases. The isochronous
communication link is first characterized experimentally for
functionality and operating margin using a test circuit with on-chip
interconnects, and then characterized over an MCM and over a 20\,mm Nb
flex tape, with two separate resonators with tunable relative phases
for driver and receiver.

\section{Isochronous receiver data link}

The isochronous board-to-board link is based on the feature that the
resonators on all boards in the system are powered by a common
high-precision clock source. In this scenario there is no relative
phase drift in the system, but there is a static and unknown phase
offset between boards due to variability e.g., in cable length. The
phase offset is established at the receiver during initialization. The
phase can be reset with reinitialization but this is not done in normal
operation. Initialization is achieved with a global training sequence,
e.g. provided by a scan chain passing though all boards. The training
data are oversampled using the four phases of the RQL AC clock, and
the phase difference is established using logical aligning block. One
control block per bus is sufficient. Currently the receiver design
resolves 90$^{\circ}$ phase granularity and has three cycles of
latency. Increasing the oversampling ratio increases phase granularity
at the expense of latency and complexity.

The isochronous receiver one-bit link shown in
Fig.~\ref{block_diagram} consists of a bandwidth-efficient driver, a
20\,$\Omega$ line, and an oversampling receiver. The
bandwidth-efficient driver is a self-resetting gate as described in
\cite{egan2021MCM}. The driver latches to produce a Gaussian waveform of
about 10 underlying SFQ pulses per bit. The driver significantly
increases signal propagation distance by reducing high frequency
analog content and the associated dispersion. The driver supports up
to 10\,Gb/s serial data rate. The 20\,$\Omega$ line is a Nb passive
transmission line (PTL) resistively terminated at the receiver. The
PTL can be implemented as a stripline on a Si substrate or cable
flexible substrate. The particular implementations used in the current
experiments are discussed below.

The isochronous oversampling receiver circuit consists of a DC-biased
block used to receive the signal at arbitrary timing, followed by an
AC-biased block used to align the incoming pulse to a particular phase
of the global clock. The DC-biased receiver is sensitive only to
positive-polarity signals. The negative-polarity trailing pulse
generated by the driver is terminated at the receiver and lost. The
receiver is DC-powered to guarantee that signal will be captured
despite arbitrary timing relative to the RQL AC clock. The receiver
consists simply of a two-junction Josephson transmission line with
35\,$\mu$A and 50\,$\mu$A critical currents as described in
\cite{egan2021MCM}.

The received signal is fed to a DC-powered four-way
splitter. Oversampling is realized by feeding these signals in
parallel to four SFQ-to-RQL converters that are AC-biased with the
four RQL clock phases. For those phases with good timing relative to
the incoming signal, the SFQ-to-RQL stage converts the positive SFQ
into a bipolar RQL signal. Based on the relative timing of the
incoming signal and the RQL phases, one or two converters will
succeed.

\begin{figure}
 \centering
 \includegraphics[width=3.4in]{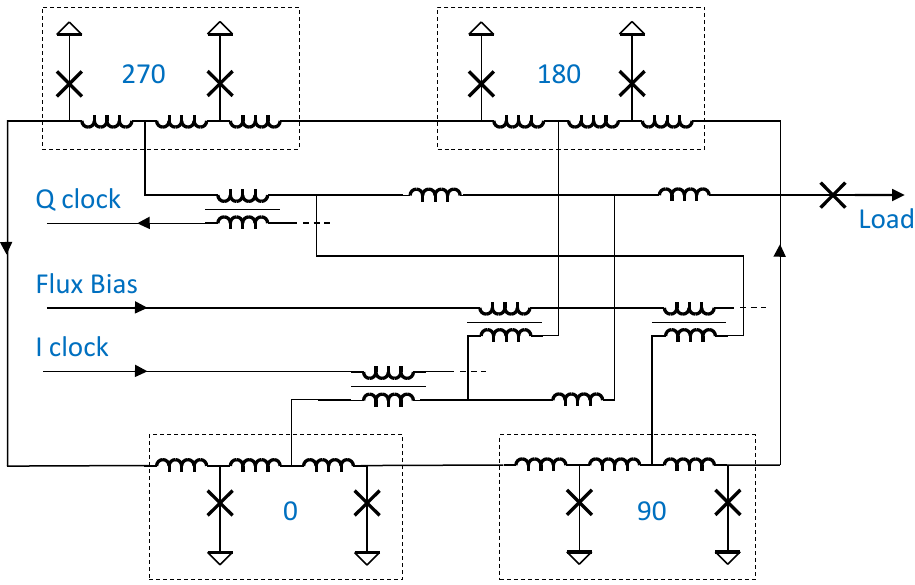}
 \caption{The DC-biased circuits are powered with a AC-to-DC converter
   called a ''flux-pump,'' consisting of a circular JTL that is
   AC-biased, but supports propagation of unipolar SFQ pulses. The
   flux bias shown produces an offset for symmetry, as with
   conventional RQL circuits. An additional flux bias (not shown)
   initializes one of the JTL signal inductors with a unipolar
   SFQ. Current regulation is provided by the junction in series with
   the load.
 \label{flux_pump}}
\end{figure}

DC bias could be supplied externally and in parallel to all receivers,
but to achieve a scalable solution we instead use a Josephson AC-to-DC
converter on-chip, called a ``flux-pump.'' The AC-to-DC converter acts
as a current-limited voltage source. The maximum voltage corresponds
to one SFQ pulse per clock cycle---just enough to power the receiver,
which consumes up to one SFQ per clock cycle from the bias line, when
fully active. The current-limit is by default a strong function of the
applied AC bias level. A floating Josephson junction in series with
the load can be used to regulate the current. A flux-pump circuit
function has been described in \cite{semenov2014new}. As a primary
figure of merit for the flux pump is high current compliance across
parameter variations, we use our own design, shown in
Fig.~\ref{flux_pump}.

\begin{figure}
 \centering
 \includegraphics[width=3.4in]{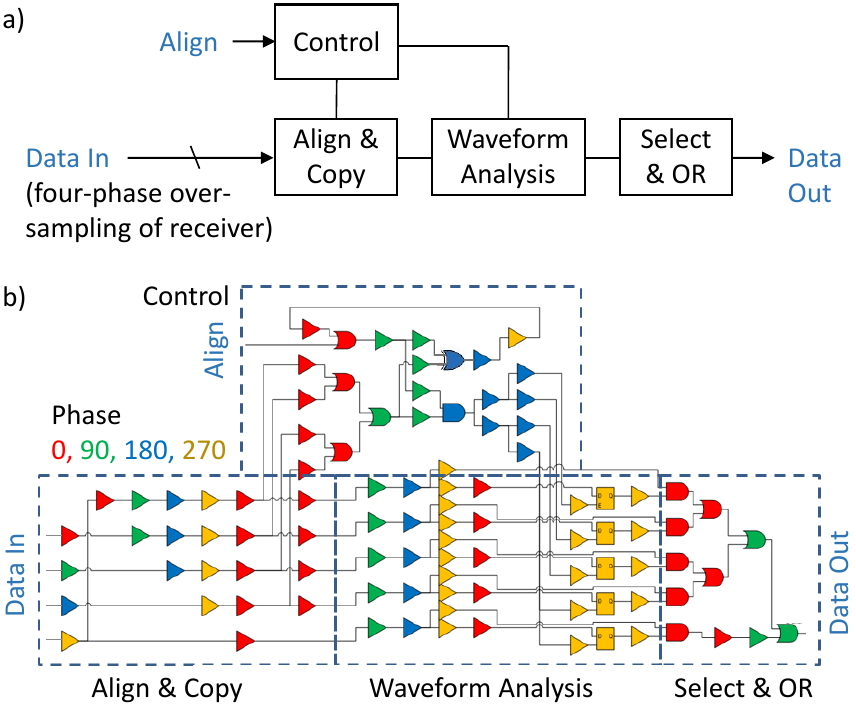}
 \caption{Isochronous receiver digital synchronization a) block
   diagram and b) schematic. The circuit detects and locks the
   relative phase between driver and receiver based on an align
   control input and data pattern received by the DC-powered
   oversampling portion of the receiver. Each triangle represents a
   two junction RQL JTL, color-coded by AC clock phase. The total
   latency through the block is three clock cycles. For this
   demonstration, the original AnotB gate has been replaced by an XOR
   gate to improve margins and show functionality of the overall
   circuit. With this schematic, unintended realignment may
   occur. Replacing XOR by an AnotB gate will fix this and will be
   used in future designs.
 \label{digital}}
\end{figure}

The digital portion of the isochronous receiver, illustrated in
Fig.~\ref{digital}, aligns the incoming signal with the local clock
phase by selecting among paths of differing delay. The different path
lengths are generated in the ``Align \& Copy'' block. One or two contiguous
paths are selected by the D-latches (squares) in the ``Waveform Analysis''
block. The ``Select \& OR'' block ANDs the incoming data with the state of
each latch, and ORs these results to produce the final output. The
incoming data constantly provides input to the D-latches, but their
state is set only when the Align signal associated with the ``Control''
block enables them. When the latches are enabled, an isolated
pilot pulse on the data input sets the latches associated with the
correct path delays, which are those that succeed in picking up the
input signal.

Five delay paths are used as the delay wraps around from 0$^{\circ}$
to 360$^{\circ}$. A signal in the gray zone between clock cycles may
be configured for minimum delay or for one additional cycle. However,
once the link is configured, the selection will be maintained for
subsequent data.

Extension to a multi-bit bus could be achieved by replicating the
receiver. If all the lines in the bus are equal in length, a common
control block associated with just one of the bits would be needed to
configure all the bits.

\section{Test vehicle}

\begin{figure}
 \includegraphics[width=3.4in]{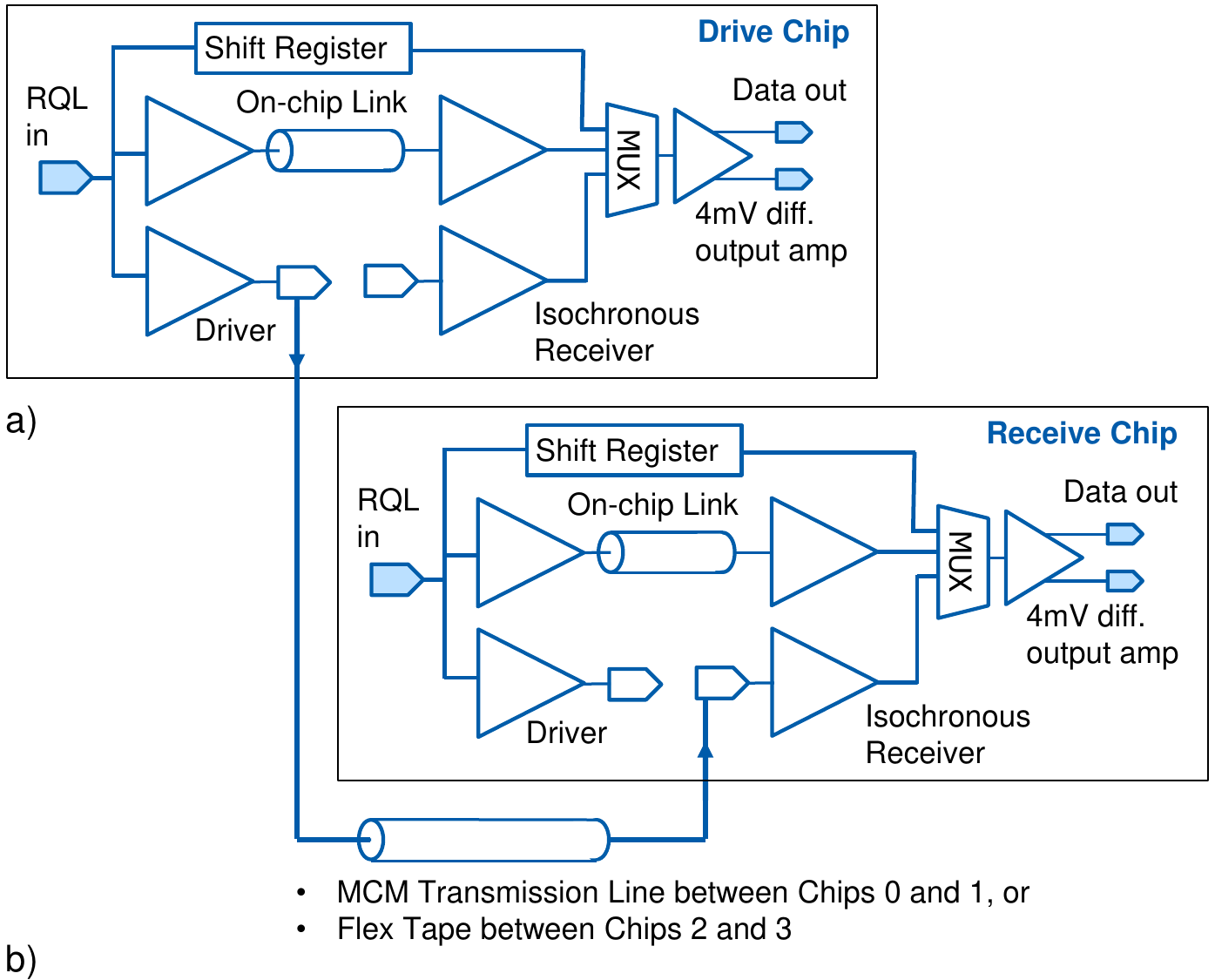}
 \includegraphics[width=2.8in]{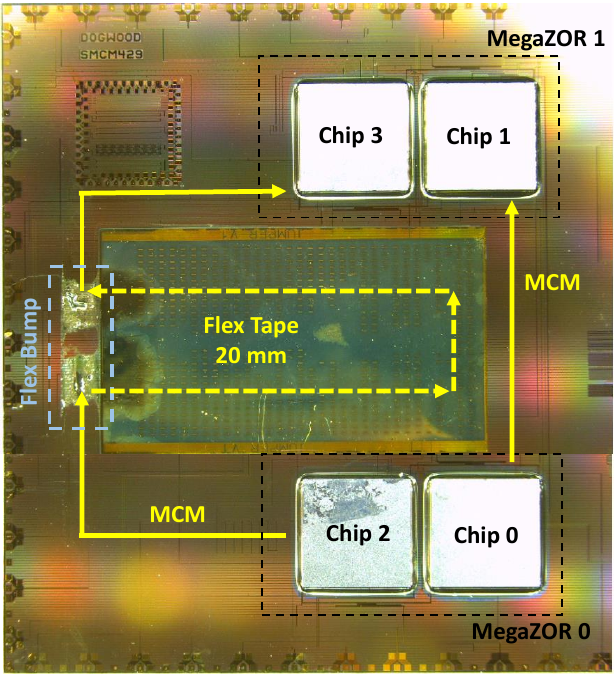}
 \caption{The isochronous data link test circuit a) block diagram,
   and b) microphotograph.  The same chip design is used for three
   experiments: on-chip PTL, MCM PTL, and flex ribbon cable. Four
   5$\times$5\,mm chips of the same design are mounted on a
   32$\times$32\,mm MCM and configured to test the MCM PTL link and
   flex link. The 10\,mm wide, 12 bit Nb flex ribbon cable is mounted
   on the MCM using a custom bump process after chip mounting. The
   four chips represent two driver-receiver pairs, with one pair for
   each link. Driver and receiver chips are on two independently
   controlled resonant clock networks, MegaZOR\,0 and MegaZOR\,1, on
   the carrier, which allows for adjustment of the relative phase
   between drive and receive.
 \label{mcm}}
\end{figure}

We have designed three experiments to test the isochronous
communication link with data transfer using on-chip PTL, on-carrier
PTL, and Nb flex cable, as shown in Fig.~\ref{mcm}a.  The on-chip data
link is for measuring the operating margins of on-chip
circuitry. The two experiments with on-carrier link and Nb flex cable
link allow to test complete functionality of the isochronous link
including phase alignment. These experiments use a multi-chip module
(MCM) with four chips mounted on superconducting carrier as shown in
Fig.~\ref{mcm}b.

\begin{figure}
 \centering
 \includegraphics[width=3.4in]{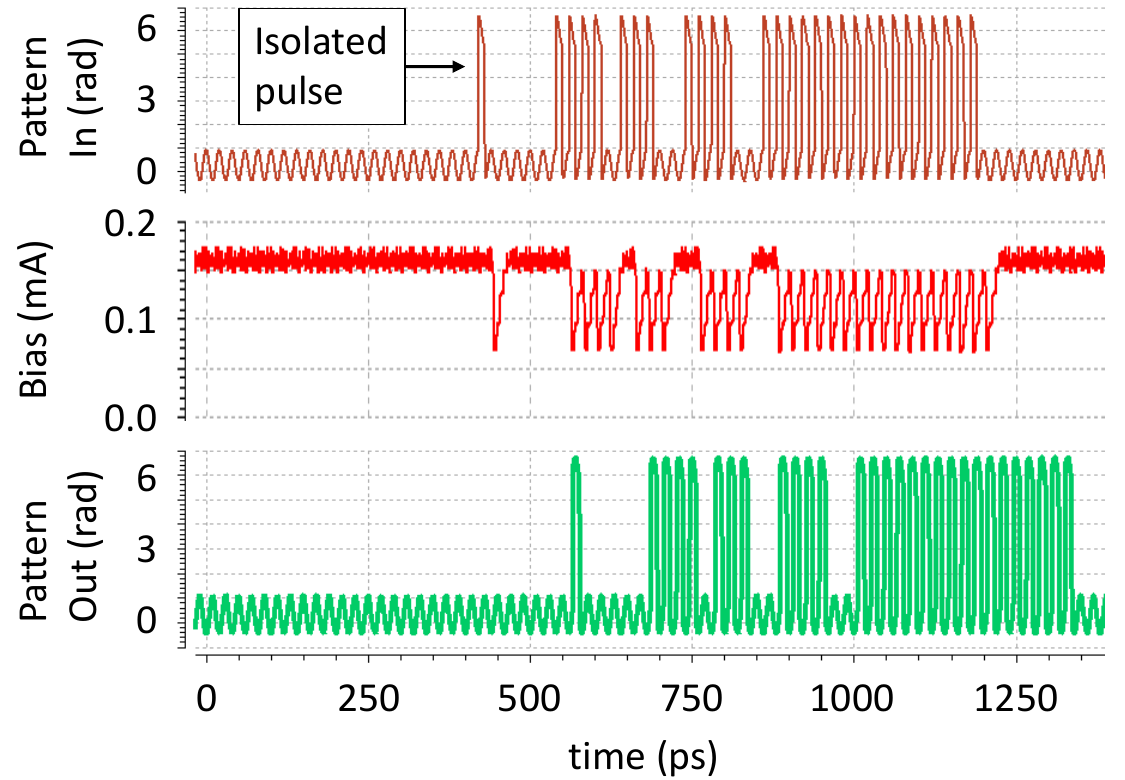}
 \caption{The input data sequence, ``Pattern In'' containing the
   isolated training pulse followed by the variable data pattern is
   sent by the driver through the transmission line and received by DC
   portion of the receiver. The flux-pump generated ``Bias'' follows the
   digital pattern producing current swing of 71-148\,$\mu$A at each
   digital one. The data is aligned using the digital block and
   replicated on ``Pattern Out.''
 \label{waves}}
\end{figure}

The on-chip circuitry has all components as described in the previous
section. The on-chip data link is configured to be tested on
individual die prior to die attach on MCM. It contains 22\,mm long,
20\,$\Omega$ on-chip transmission line with propagation delay
targeting a gray zone between 0$^{\circ}$ and 270$^{\circ}$
phase. This case represents the most complicated regime of phase
alignment resolving the phase difference at the boundary between two
clocks. The data link circuit is wrapped with the on-chip test bed
(see Fig.~\ref{mcm}a) consisting of two differential inputs
\cite{herr2011ultra} for data and align signals and a single 4\,mV
differential output amplifier \cite{egan2021true}. The whole circuit
is powered by an on-chip resonant clock network covering the active
area of the chip. The circuit has common AC bias and DC flux lines for
all components. The test circuit was designed to operate at 3.4\,GHz,
which is lower than the 10\,GHz target for the design of the data link
itself. The lower operating frequency allows feasibility demonstration
using the simple fabrication process and design flow that was not
targeted to aggressive timing. Simulation waveforms are shown in
Fig.~\ref{waves}. The simulated circuit operating margin for AC bias
is 5\,dB ($\pm$30\%) with junction critical current margins of
individual gates at $\pm$50\%. The isochronous receiver contains about
600 Josephson junctions with total power dissipation less than
5\,fJ per bit including all receiver overhead and a cryocooler
efficiency of 300\,W/W.

The chip was designed into the fabrication process at D-Wave with
0.25\,$\mu$m feature size and six metal layers
\cite{berkley2010scalable}, and with a modified critical current
density of 100\,$\mu$A/$\mu$m$^2$. The minimum critical current in the
circuit is 35\,$\mu$A.  The analog portion of the circuit consisting
of the driver, flux pumps and DC receiver, was designed using a custom
design flow in the Cadence environment based on Spectre analog
simulations and inductor P-cell based layout. The digital block was
designed using automated Cadence based digital flow including
synthesis, placement, timing and logical verification. A custom
inductance-target router was used. Final chip-level verification
includes Spectre simulations as well as DRC and LVS with custom checks
for parasitic coupling.

For the MCM phase alignment experiments we used the well-characterized
process for the carrier and bump-bonds developed at Lincoln Laboratory
\cite{LincolnMCM}. The ``smcm4m'' four Nb metal
process supports the design of both 20\,$\Omega$ data PTLs and the
clock resonator network, isolated using a superconducting ground
plane. The In bump bonding process, with 15\,$\mu$m bump
diameter, 35\,$\mu$m bump pitch, and about 3\,$\mu$m bump height
post-bonding, supports MCM-to-chip transitions with 350\,GHz analog
bandwidth \cite{egan2021MCM}.

\begin{figure}
 \centering
 \includegraphics[width=2.8in]{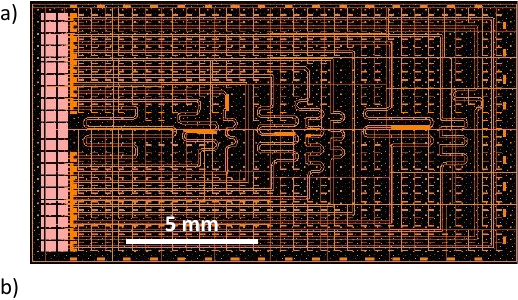}
 \includegraphics[width=3.4in]{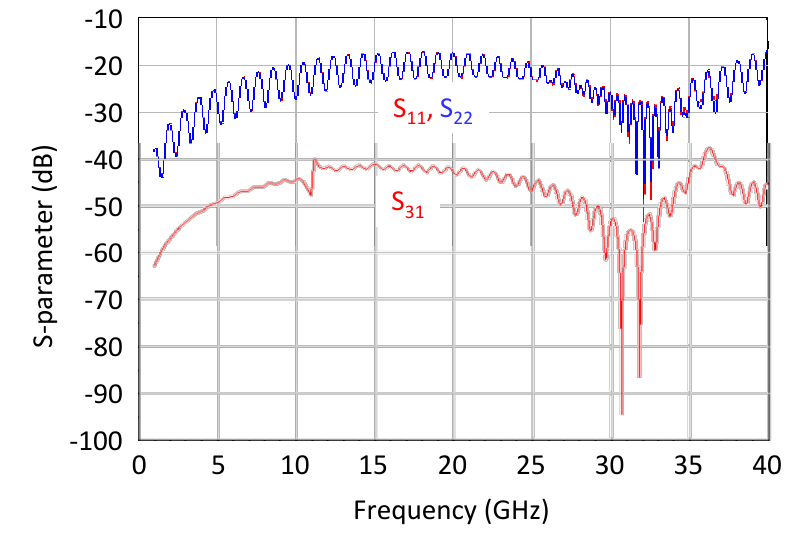}
 \caption{The flex cable supports three 4-bit matched length buses
   that loop back to the single connection. Only one trace is used in
   the current MCM experiment with all other traces connected to MCM
   ground. a) Physical layout. b) Simulated S-parameters using an HFSS
   3D EM model for the link consisting of the MCM-to-cable transitions
   and the 20\,mm flex cable. S$_{11}$ and S$_{22}$ are the return
   loss, and S$_{31}$ is the parasitic coupling to an adjacent line.
 \label{tape}}
\end{figure}

The 32$\times$32\,mm MCM carrier for the phase align experiments with
both on-carrier PTL and Nb flex cable, contains two separate
``megaZOR'' resonator networks, with independent control of frequency
and amplitude. Each MegaZOR is a resonator of $\lambda/2$ segments on
the MCM designed to drive several on–chip ZOR resonators as described
in \cite{egan2021MCM}. The carrier MegaZOR design has been targeted to
match the experimental average resonance frequency of the individual
chips. The MegaZOR was optimized to achieve less than 1\,dB amplitude
variation and less than 5$^{\circ}$ phase variation between two chips
with worst-case $\pm 3$\% speed of light variation across the carrier
and $\pm 1$\% variation in resonance frequency between individual
chips.

The superconducting flexible ribbon cable has been designed at Auburn
University \cite{tuckerman2016flexible, gupta2019thin} and fabricated
by Hightec, a commercial manufacturer. The cable is a stack-up of
thin-film polyimide layers (PI2611 from HD Microsystems) and patterned
Nb metal. There are three Nb metal layers building 86\,$\mu$m-wide
stripline interconnects with high isolation. The signal traces were
designed by Auburn to be 20\,$\Omega$, matching the other link
components. For the current experiment, the custom tape has all wires
in loop-back as shown in Fig.~\ref{tape}a.

The transition between MCM and flex tape is done using a second In
bump-bond process with 75\,$\mu$m bumps at 150\,$\mu$m pitch
\cite{peek1fabrication}. The transition between cable and MCM has been
optimized using HFSS 3D field solver with requirements of less than
$-20$\,dB reflection and less than $-40$\,dB coupling between adjacent
pins, as shown in Fig.~\ref{tape}b. Full system simulations in Spectre
have been done to verify circuit operating margin using the
S-parameter matrix for the MCM-flex transition extracted from HFSS
model. The operating margin for the circuit is comparable to on-chip
link simulations.

\section{Test results}

Candidate chips from the same wafer were characterized individually for
parameter targeting using PCM data and for resonator performance
following procedures explained in \cite{strong2021ZOR}. The PCM
measurements indicated that the Josephson junction critical current
and routing inductances are within tolerances of $\pm$10\% and
$\pm$2\% respectively. The average wafer resonant frequency of
3.6\,GHz is within 6\% of the 3.4\,GHz design target. S-parameters for
the resonator showed less than $-$40\,dB coupling and coincident
resonant peaks between I and Q resonators.

\begin{figure}
 \centering
 \includegraphics[width=3.4in]{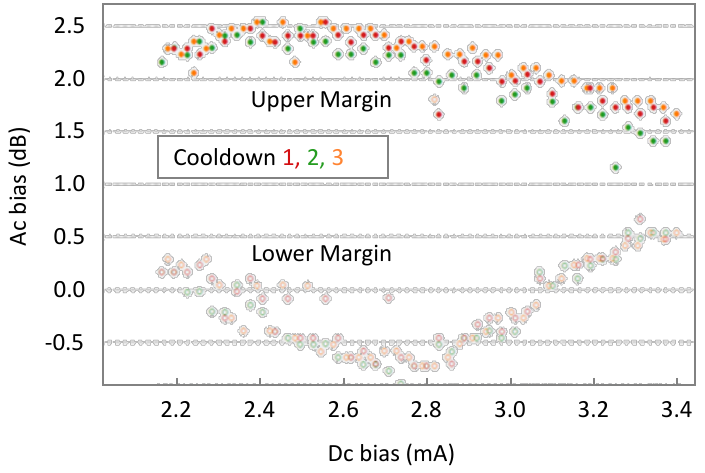}
 \caption{On-chip isochronous link.  2D operating margin of AC power
   vs.\ DC flux bias measured across three cooldowns. Operating margin
   exhibits the typical-for-RQL V-shaped operating region on the low
   end limited by positive and negative pulse annihilation. The
   high-end margin is limited by overbias of the Josephson
   junctions. The reported margin of 3\,dB corresponds to the
   inflection point of lower margin.
 \label{2d}}
\end{figure}

The selected chips were tested for functionality of the on-chip data
link. The test was performed at the resonance frequency in liquid He
using a pressure-contact dip probe and a custom electronics rack. The
digital vectors were generated using Keysight M8020 synchronized with
a Rhode \& Schwartz SGS100A frequency generator. The output data link
was as described in \cite{egan2021true}. The output signal was captured on
a Tektronix MSO7204C oscilloscope, and an automated procedure for
collecting 2D margins of AC clock amplitude vs.\ JTL bias was applied.

\begin{figure}
 \centering
 \includegraphics[width=3.4in]{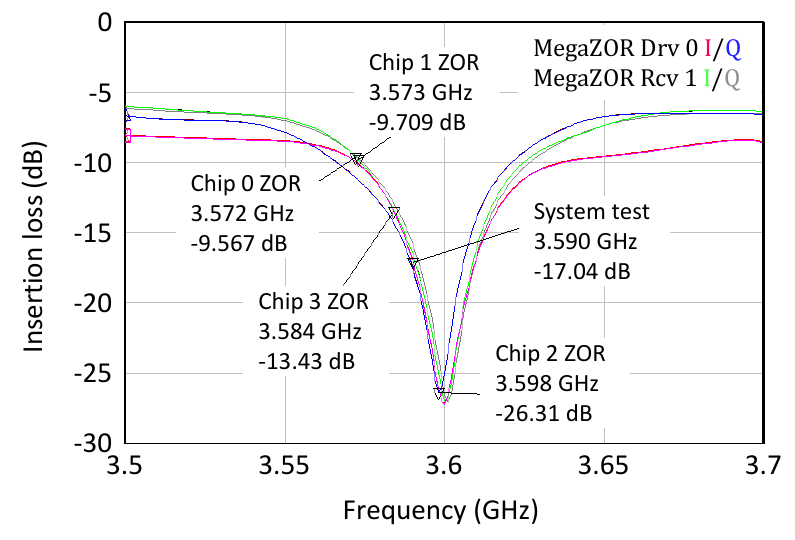}
 \caption{Measured insertion loss of the MCM with four bonded
   chips. Individual chip resonant frequencies are marked on top of
   four curves corresponding to I and Q resonators of two MegaZORs on
   the carrier. MegaZOR\,0 consisting of Chips 0 and 2 has a resonance
   of 3.598\,GHz, and MegaZOR 1 consisting of Chips 1 and 3 has a
   resonance of 3.60\,GHz, System test was performed at 3.59\,GHz.
 \label{s-zor}}
\end{figure}

Fig.~\ref{2d} shows experimental 2D operating margin on the bias. The
test pattern starts with align input bit and data alignment pattern
with an isolated bit, followed by nine zeros. The data test pattern
consists of four ones followed by four zeros and 16 ones. The same
test pattern is repeated for each point for 2D margin collection. The
chip has about 3\,dB AC clock margin at optimum JTL bias of
2.75\,mA. This chip exhibits 0.5\,dB variation across cooldowns, which
is within design tolerances for coupling to parasitic flux vortexes
sequestered in the moats. Margin degradation of 2\,dB relative to
simulations is within the typical budget for the fabrication process
and layout style, with contributions including variation in fabrication
parameters, resonator uniformity, and cross-net coupling in the digital
routing.

The carrier was selected from a wafer with 16 32$\times$32\,mm
instances. Each carrier was characterized using on-carrier PCMs
containing a chain of the MCM bumps for DC measurements of critical
current, and a $\lambda/2$ resonator for speed-of-light measurements
for data and resonator metal layers.  PCM results showed 100\% yield
on DC bump chains with more than 32\,mA critical current and about 3\%
$\lambda/2$ resonance frequency variation across the wafer with the
best die being within 1\% of the design target. Test results of
S-parameter characterization of each MegaZOR are shown in
Fig.~\ref{s-zor}.

\begin{figure}
 \centering \includegraphics[width=3.1in]{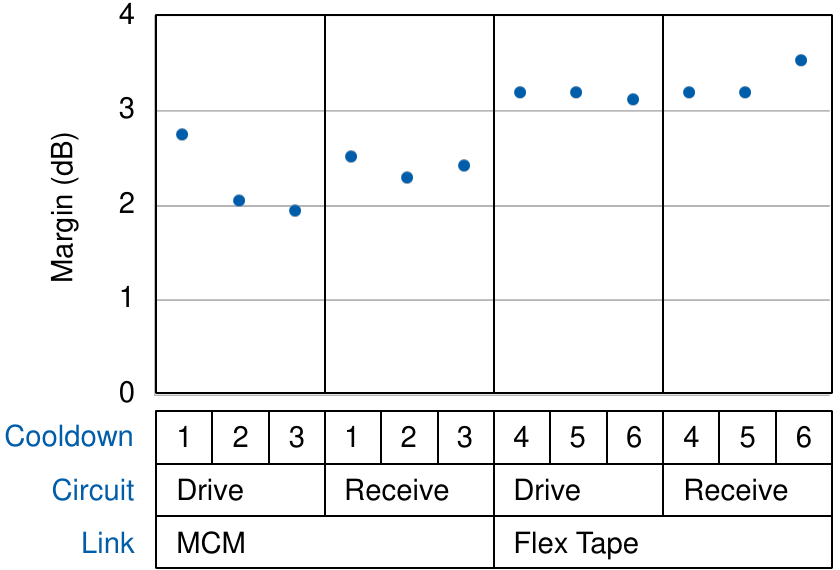}
 \caption{Operating margin of MCM data links measured for both
   on-carrier PTL and Nb flex cable across three
   cooldowns. Experiments use the same data pattern and test protocol
   for both links. AC power margins on the driver are measured while
   holding the receiver at nominal, and the driver is held at nominal
   while measuring the receiver.
 \label{marg}}
\end{figure}

Fig.~\ref{marg} shows experimentally-measured margins for two data
links on the MCM configured to 90$^{\circ}$ static phase offset
between driver and receiver. The same test pattern has been used as
for the individual on-chip link. Both links showed complete
functionality with operating margin of 2.8\,dB on-carrier and 3.5\,dB
on the flex cable. Operating margins for both links are on par with
on-chip PTL test and stable across cool downs. The best margin of
3.5\,dB is for the data link over the flex cable, which can be
accounted for by low-loss in the polyimide dielectric compared to
SiO$_2$ dielectric on the carrier, and by the wider geometry of the
flex Nb signal lines with less dispersion.

\section{Conclusion}
We have presented the first practical circuit for superconducting
board-to-board communication, based on an isochronous data link
enabled by the resonant clock of RQL circuits. The data link was
demonstrated with a 20\,mm Nb flex tape mounted on a superconducting
MCM. The 3\,dB margin of the flex-tape data link was comparable to a
similar circuit using superconducting PTLs on the MCM carrier.

The isochronous link explained in this paper is scalable to multi-bit
buses and higher frequency. More optimal phase selection at the
receiver could be achieved by adding more granularity in phases while
leaving the overall design unchanged. For a low skew, multi-bit bus,
just one of the bits is required for phase selection, and its digital
control block can be shared with the other bits. Where skew between
bits exceeds clock phase granularity, the entire digital block needs
to be duplicated per bit, or per group. The analog bandwidth of the
link is limited by the dispersion and loss in the flex cable. The
driver produces data pulses of about 10 SFQ pulses per bit, allowing a
data rate of 10\,Gb/s, and a flex cable length up to two meters.

Small hardware overhead, energy efficiency, and high throughput of the
isochronous link positions RQL technology as a digital technology with
high computational density and high cross-sectional bandwidth of the
interconnect in a system of multiple boards. In particular, the
interconnect could enable flat memory access across the system. Such
an architecture has the potential for significant performance gains in
systolic arrays for machine learning.

\begin{acknowledgments}

The authors acknowledge the contributions of the Lincoln Laboratory
team, particularly Sergey Tolpygo, in technical discussions and
assistance with design. We also acknowledge Vaibhav Gupta, Archit Shah
and the AMNSTC staff (especially Drew Sellers) at Auburn University
for their assistance with fabrication of the flex cables. The Northrop
Grumman superconducting digital EDA and PDK teams laid the foundation
for digital design of the isochronous receiver. We also acknowledge
the contribution of Steve Shauck and Alex Braun for guidance with the
digital design.

\end{acknowledgments}

\bibliography{isoc}

\end{document}